\documentclass[aps,pra,twocolumn,showpacs,amsmath,amssymb]{revtex4}
\usepackage{graphicx}
\usepackage{dcolumn}
\usepackage{bm}
\begin{document}

\title{Analysis of an atom laser based on the spatial control 
of the scattering length.}
\author{Alicia V. Carpentier, Humberto Michinel, 
and Mar\'{\i}a I. Rodas-Verde}
\affiliation{\'Area de \'Optica, Facultade de Ciencias de Ourense,\\ 
Universidade de Vigo, As Lagoas s/n, Ourense, ES-32004 Spain.}

\author{V\'{\i}ctor M. P\'erez-Garc\'{\i}a}
\affiliation{Departamento de Matem\'aticas, E. T. S. I. Industriales,\\ 
Universidad de Castilla-La Mancha, 13071 Ciudad Real, Spain.}

\begin{abstract}
In this paper we  analyze atom lasers based on the spatial modulation 
of the scattering length of a Bose-Einstein Condensate. We demonstrate, through 
numerical simulations and approximate analytical methods, the controllable 
emission of matter-wave bursts and study the dependence of the process on the 
spatial shape of the scattering length along the axis of emission. 
We also study the role of an additional modulation of the scattering length in time.
\end{abstract}


\maketitle
\section{Introduction}

Atom lasers are sources of coherent matter waves that  
use an ultra-cold gas of trapped alkalii atoms as a reservoir
from which coherent pulses of thousands of atoms are extracted.  Since the 
experimental achievement of Bose-Einstein Condensation 
(BEC) in gases \cite{Anderson95}, several 
methods have been proposed to deliver atoms from their confinement. 
The first of these devices used short radio-frequency pulses as 
outcoupling mechanism, flipping the spins of some of the atoms to
release them from the trap \cite{Mewes97}. Later, other atom lasers 
were built leading to pulsed, semi-continuous or single-atom coherent 
sources \cite{Bloch,Hagley,Bloch2,Pepe,Ultimo}.

The quality of an atom laser is given by the amount of atoms that can 
be delivered from the trap and by the purity of the emission process.
Concerning this point, it has been recently shown \cite{close} that
spin-flipping techniques present serious limitations in the number of
atoms that can be emitted. This effect is related with the fluctuations 
at high flux due to the fact that the output-coupling mechanism populates all 
accessible Zeeman states. This constitutes a strong drawback for practical 
applications of atom lasers in high precision measurements like matter 
wave gyroscopes \cite{gustavson97}.

On the other hand, an atomic soliton laser using the mechanism of 
modulational instability was proposed in Ref. \cite{Carr04}. In this case,
the emission is obtained by the combination of purely nonlinear effects in the 
atom cloud and the relaxation of the trap, it being necessary that the
total number of particles in the cloud exceeds a critical threshold. This system 
has the advantage over standard atom lasers of producing matter wave pulses in the form of 
{\em solitons} \cite{Perezgarcia98, solitons1,solitons2}, a kind of nonlinear waves generated
by a perfect balance between dispersive and nonlinear effects, yielding 
to robust wave packets that propagate without shape distortion \cite{zakharov}. 
Other types of atom laser based on nonlinear effects have been proposed \cite{Chen}.

However, although one could extract a few coherent solitons from a 
Bose-Einstein Condensate by the mechanism of modulational instability, the final output would be 
very limited since the  number of atoms per pulse generated by this phenomenon is 
only a small fraction of the initial number of atoms in the condensate 
due to collapse processes. Moreover, the number of solitons generated is not 
large and half of them would be directed backwards. Finally,  the trap must
be destroyed for outcoupling and the pulses travel at different 
speeds once the trap is removed. Thus, it is important to discuss new 
outcoupling mechanisms for atom lasers. This is specially interesting 
since the techniques for generating BECs with large numbers of particles 
and their physical properties are nowadays well established and the 
current challenges in the field face the design of practical devices \cite{chip}.

In a recent work \cite{rodas-verde05} a novel outcoupling mechanism  
for an atom laser has been proposed. The method is based on the fact that
a spatial variation of the scattering length ($a$) (see Fig. \ref{fig1}) 
can be used to extract a controllable train 
of up to several hundreds of atomic solitons from a BEC without altering 
the trap properties. In Ref. \cite{rodas-verde05}, a simple model for the 
spatial variation of the scattering length (a step function) was used in order 
to introduce the basic ideas and illustrate the phenomenon. 
However, for the practical realization of such device, a deeper analysis must be done, including 
studying the role of more physical distributions of the spatial dependence of $a$, 
and the temporal control of the output by the considering an additional time dependence 
of the scattering length. In this paper we consider 
these aspects of the soliton emission process which are relevant for the theoretical 
understanding and experimental demonstration of this new type of atom laser.

The structure of this paper is as follows: in Section \ref{II}, we
present the configuration of the system and the mathematical model to be 
used for the theoretical analysis. In Section \ref{III} we study the mechanism 
of emission of single solitons. To do so we first study a simplified model 
for which an analytical study can be done. We also analyze theoretically a 
more realistic scheme by means of variational methods and numerical simulations,  
for Gaussian-shaped spatial variations of the scattering length. Finally, in Section \ref{IV} we
study the effect of temporal variations of the scattering length induced by 
pulsed beams. 


\section{System configuration and theoretical model}
\label{II}

\begin{figure}[htb]
{\centering \resizebox*{1\columnwidth}{!}{\includegraphics{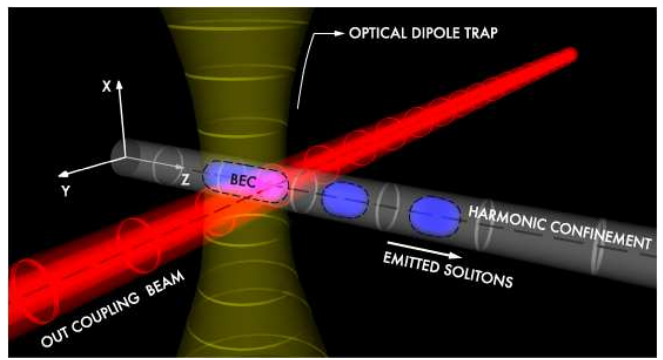}} \par}
\caption{[Color online] Sketch of the system we will study in this paper 
for the case of optically 
controlled scattering length showing the BEC in the optical dipole trap, 
the transverse magnetic confinement  and the laser beam used to manage the scattering length region.
For a critical number of atoms or equivalently below a critical value of the scattering length, a burst of matter-wave solitons 
is emitted along the weakly confining axis.}
\label{fig1}
\end{figure}

As in Ref. \cite{rodas-verde05}, we assume a BEC which is 
strongly confined in the transverse directions ($x,y$) and weakly along the longitudinal 
one ($z$) leading to a {\em cigar-shaped} configuration.
We will consider the effect of a spatial variation of the scattering length
along $z$ from positive (or zero) to negative values. In principle this can be 
done by magnetic \cite{FB1} or optical \cite{FB2} means, for instance by using an appropriate
laser beam that shines one of the edges of the condensate as in  the sketch of the system 
plotted in Fig.\ref{fig1}. In this case the region of negative scattering length can be 
varied and displaced along the condensate by simply moving the laser beam. The 
variation of $a$ can also be controlled in time by using pulsed laser beams. Similar arguments
apply for magnetically controlled scattering length, although the control of laser allows for 
faster and easier manipulation of the spatial variations of the scattering length. In any case 
what it is important is that we will depart from the model of step-like spatial variation 
of the scattering of Ref. \cite{rodas-verde05} to more realistically 
achievable smoother dependences.

When the zone in which the scattering length is managed to negative values
overlaps with the wings of the atom cloud, as in  Fig.\ref{fig1}, the BEC 
feels a repulsive force whose strength depends on the number 
of atoms in the condensate. When this effect is large enough, the 
trap is overcome and part of the cloud is delivered and emitted 
outwards. When the condensate refills the gap left out by the outgoing pulse 
the process starts again and a new soliton is emitted. This process 
would continue while there is a large enough remnant of atoms in the trap 
and would lead to a soliton burst escaping from the BEC.

Therefore, we will consider that the cloud of $N$ equal bosons of mass $m$ 
is tightly trapped in ($x,y$) by a harmonic potential $V_\perp$ of frequency
$\nu_\perp$ and weakly confined along $z$ by the effect of an optical dipole trap 
$V_z$\cite{Stamper98,Martikainen99} that can be produced by a laser beam 
of a given width along $z$ (see Fig. \ref{fig1}). Thus, we have
\begin{multline}
V(\vec{r})=V_\perp+V_z= \\
\frac{m\nu^2_\perp}{2} 
\left( x^2+y^2 \right)+V_0
\left[1-\exp\left(-\frac{z^2}{L^2} \right)\right],
\end{multline}
where $V_0$ is the depth of the shallow optical dipole potential and $L$ its
characteristic width along $z$. The choice of a shallow Gaussian trap is 
very important, since the goal is to obtain outcoupling of solitons along 
the $z$ axis. This can be achieved with a potential barrier that can 
be overcome by the self-interaction effects. The dynamics of the previous
system in the mean field limit is described by a Gross-Pitaevskii equation 
of the form:
\begin{equation}
\label{GPE}
i \hbar \frac{\partial \Psi}{\partial t} =
- \frac{\hbar^2}{2 m} \nabla^{2}\Psi +
V(\vec{r})\Psi + U(z)|\Psi|^2 \Psi,
\end{equation}
where $\Psi$ is the condensate wavefunction, and its norm
$N = \int |\Psi|^2 \ d^3 \mathbf{r}$ gives the number of particles.
The coefficient $U(z) = 4 \pi \hbar^2 a(z)/m$ depends on the 
scattering length $a$, which characterizes the 2-body interactions
between atoms. As commented previously, we will consider $a$ 
to be a function of $z$ with a localized region in which it becomes negative.

We will concentrate on situations in which the spatial size of the ground 
state of the optical dipole trap is much larger than the ground state of the 
transverse harmonic potential leading to effectively one-dimensional dynamics. 
In this situation we can describe the dynamics 
of the condensate in the quasi-one dimensional
limit as given by a factorized wavefunction of the form \cite{Perezgarcia98} 
$\Psi(\boldsymbol{r},t)=\Phi_0(x,y)\cdot\psi(z,t)$, satisfying
\begin{equation}
\label{NLSE}
i\frac{\partial \psi}{\partial \tau} = - \frac{r_{\perp}^2}{2}
\frac{\partial^2\psi}{\partial z^2} + f(z)\psi + g |\psi|^2 \psi,
\end{equation}
where $r_\perp = \sqrt{\hbar/m\nu_\perp}$ is the transverse size of the cloud,
 $f(\eta) = V_z/(\hbar\nu_\perp)$, $\tau=\nu_\perp t$ is the time measured 
in units of the inverse of the radial trapping frequency
and $g(z) =2\pi r_\perp^2 a(z)$ is the effective interaction 
coefficient. 

As an example we will present specific numbers in this paper corresponding to 
 $^7$Li, using the experimental parameters of Ref. \cite{solitons1}
$V_0=\hbar\nu_\perp/2$, $\nu_\perp=1KHz$, $L=4r_\perp$, $N=3\cdot10^5$, $w=5.4r_\perp$ 
$a=-1.4nm$ and times ranging from $t=0$ to $t=1s$. However, our results hold 
for different atomic species like $^{85}$Rb and $^{133}$Cs, with adequate parameters. 

\begin{figure}[htb]
{\centering \resizebox*{1\columnwidth}{!}{\includegraphics{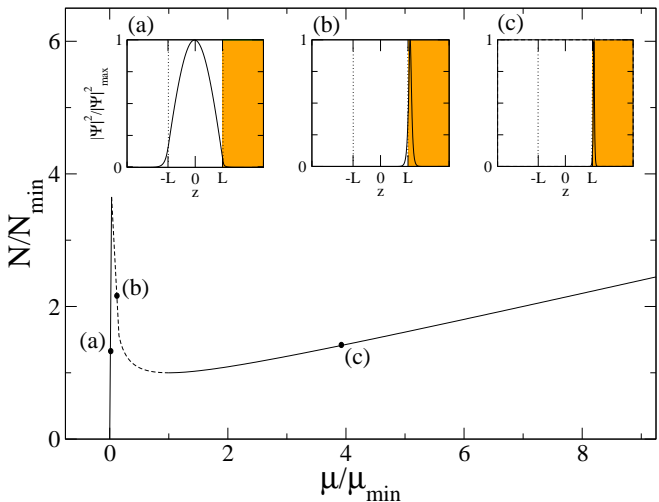}} \par}
\caption{[Color online] Dependence of the number of particles in the condensate 
$N$ on the chemical potential $\mu$ for the simple case of square-shaped potential. The 
values are normalized by $N_{min}$ and $\mu_{min}$ which correspond to the minimum 
of the curve. For the experimental values considered in the text
for $^7$Li we find $N_{min}=2\times10^4$ and $\mu_{min}=\hbar\nu_\perp/2$ 
for $a=-0.3$ nm. The dotted part of the curve $\mu(N)$ corresponds to an 
unstable region. The insets show the eigenmode profiles for different values 
of $\mu$, the dotted lines indicating the boundaries of the trap. The region 
of negative scattering length is the shaded one.}
\label{fig2}
\end{figure}
                     

\section{Single soliton emission}
\label{III}

\subsection{Exact theory for square potentials}

As we have pointed before, we will consider a situation in which  
the scattering length is changed on a localized region near the
edge of the optical trap. In that case, it is possible to understand the basics of the 
phenomenon 
of the emission of a single soliton using a simple model. We consider a 
 square trapping potential of width $L$ and depth $V_0$ and  
 a step-like scattering length, as it is illustrated in 
Fig. \ref{fig2}. Mathematically
\begin{subequations}
 \begin{equation}
\label{toy}
V(z) = \begin{cases} 0 & z< -L\\
V_0 < 0 & -L< z< L\\
0 & z> L 
\end{cases}
\end{equation}
and
\begin{equation}
a(z) = \begin{cases} 
0 & z< L\\
a_*<0 & z> L 
\end{cases}
\end{equation}
\end{subequations}
Although this model is a rough simplification of the more realistic situations to be described in detail later on, it captures the big picture
of the process. In fact, this model 
can be solved analytically, by simply calculating the solution of the 
Schr\"odinger equation in the three zones. Defining
$\gamma_1=2\mu/r_\perp\hbar\nu_\perp$, $\gamma_2=2(\mu-V_0)/r_\perp\hbar\nu_\perp$ 
and $\gamma_3=2\mu/r_\perp\hbar\nu_\perp$, we get
 \begin{equation}
\label{modofundamental}
\Psi(z) = \begin{cases} A_1 \ \text{exp}(-\gamma_1z) & z< -L\\
A_2 \ \text{cos}[(z-z_2)\gamma_2] & -L< z< L\\
A_3 \ \text{sech}[(z-z_3)\gamma_3] & z> L 
\end{cases}
\end{equation}
As it can be appreciated in the insets of Fig. \ref{fig2}, the shape of
the wavefunction depends on $a_*$. Thus, a potential connected to a region where atom-atom 
interactions are activated, displays a continuum of stationary 
fundamental states. For values of $a_*$ close to zero, (or equivalently a 
low number of atoms $N$ in the BEC), the cloud is located at the
center of the trap. As the product $Na_*$ is increased, the center of the
cloud is displaced towards the boundary between zones two and three. For
higher values of $Na_*$, the cloud is completely located in the region
with $a<0$ and takes the form of a soliton with a hyperbolic secant profile.
The continuity of the wavefunction of the condensate and its derivatives 
at the boundaries between the three regions yields to the 
following relationships:
\begin{subequations}
\begin{eqnarray}
\label{constantes}
z_2 & = & -\left[L+\frac{1}{\gamma_2} \ \tan^{-1}{\left(\frac{-\gamma_1}{\gamma_2}\right)}\right]\\
z_3 & = & L-\frac{1}{\gamma_3} \ \tanh^{-1}{\left[\frac{\gamma_2}{\gamma_3} \ \tan{\left[\gamma_2(L-z_2)\right]}\right]}\\
A_1 & = &  A_3 \ \text{sech}[\gamma_3(L-z_3)] \ \sec{[\gamma_2(L-z_2)]} \\
A_2 & = & A_1 \ \cos{[-\gamma_2(L+z_2)]} \ e^{\gamma_1L}\\
A_3 & = & \frac{1}{\text{sech}z}\sqrt{\frac{2 \text{sech}^2z-\frac{2\mu}{r_\perp^2\hbar\nu_\perp}+1}{\sqrt{8}\pi r_\perp a_*}}.
\end{eqnarray}
\end{subequations}
The number of atoms $N$ in the stationary state can be easily calculated
by integrating $|\psi|^2$ over $z$,
\begin{multline}
\label{potencia}
N  = \frac{A_2^2}{2\gamma_1} \ \text{exp}(-2\gamma_1L) +\frac{A_3^2}{\gamma_3}\left\{1-\tanh{[\gamma_3(L-z_3)]}\right\}  \\
 + \frac{A_1^2}{2}\left[2L+\frac{\sin{(2\gamma_2L)}\cos{(2\gamma_2z_2)}}{\gamma_2}\right].
\end{multline}
The previous equation provides the dependence of the chemical potential of the ground state on the 
number of atoms. As it can be seen in Fig. \ref{fig2} 
the curve $N$ vs $\mu$ for all the stationary ground states
has one maximum and one minimum, and displays a negative slope 
between them. It is well known from the theory of nonlinear 
Schr\"odinger equations \cite{litvak,akhmediev} that the stationary
states corresponding to the
zone with $dN/d\mu<0$ are unstable, what is known as the 
Vakhitov-Kolokolov's criterium of stability.
This fact implies that any small perturbation will affect dramatically the eigenstates in this 
zone, that are basically nonlinear surface waves located in the 
boundary between the linear potential and the nonlinear region. 
As a consequence of this instability, the surface wave is reflected 
by the boundary and emitted to the nonlinear zone.


\subsection{Approximate theory for realistic potentials}

The analysis of the previous section, although exact, is only a crude representation of the real potentials which
can be used in experimental scenarios. 
To study more physical spatial distributions of $a$ we will 
combine the use of numerical methods with approximate averaged Lagrangian formalisms \cite{AL} consisting on  
minimizing the Lagrangian density over a family of trial wavefunctions which we will choose as
\begin{equation}
\psi(z,\tau)=A\exp{\left[-\frac{\left(z-z_0(\tau)\right)^2}{2w^2}\right]}
\exp{\left[iv(\tau)z\right]},
\end{equation}
This ansatz leaves a single free degree of freedom for the atom cloud: its center
$z_0(\tau)$, which moves with speed proportional to $v(\tau)$.

To study an scenario close to a possible experimental setup in which the scattering 
length is optically managed, we will analyze the soliton emission when the laser beam used to manage the scattering length takes the form
\begin{equation}
\label{gaus}
a(z) =  \begin{cases} a \exp\left[-\left(\frac{z-D}{w_g}\right)^m\right],
 & z< D+w_g \\
a , & z>D+w_g,
\end{cases}
\end{equation}
i. e. an hypergaussian ramp connected at its maximum to an infinite plateau.
This choice allows us to cover many cases between the two limits $m=2$ corresponding to a Gaussian 
distribution and $m\rightarrow\infty$ representing a step function. The standard 
calculations taking $a(z)$ as given by Eq. (\ref{gaus}) lead to a Newton-type equation of the form
\begin{subequations}
\begin{equation}
\label{variacional}
\ddot{z_0} = -\frac{d\Pi}{dz_0},
\end{equation}
where $\Pi(z_0)$ is given by
\begin{multline}
\label{potential}
\Pi(z_0)  =  r_\perp^2\left[\frac{V_0}{\hbar\nu_\perp}
\left(1- \frac{e^{-\frac{z_0^2}{w^2+L^2}}}{\sqrt{1+\frac{w^2}{L^2}}}\right) + \right .  \\ \left .+\frac{N}{2w^2\pi}\int_{-\infty}^{\infty} \text{exp}\left[\frac{-2\left(z-z_0\right)}{w^2}\right]a(z) dz\right],
\end{multline}
\end{subequations}
From Eqs. \eqref{variacional} we see that the center of gravity of the cloud ($z_0$) 
behaves like a classical particle under the effect of a potential 
$\Pi(z_0)$. 
This provides a qualitative understanding of the soliton 
emission: for the linear case 
($a=0$) the center of the cloud is located at the bottom of the Gaussian 
trap and approximates its fundamental eigenstate [Fig. \ref{fig2} (a)]; 
as $a$ takes more negative values the effective trapping of the cloud 
is deformed and the minimum of the equivalent potential moves to the 
region with $z>0$. 

\begin{figure}[htb]
{\centering \resizebox*{1\columnwidth}{!}{\includegraphics{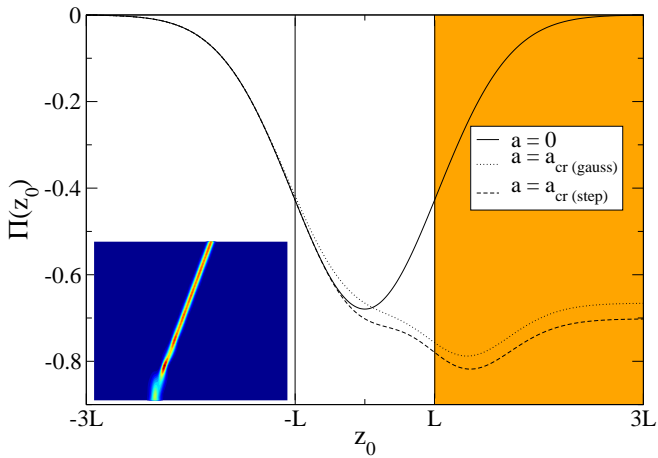}} \par}
\caption{[Color online] Effective potentials obtained for two different shapes of $a(z)$: a 
step function and a Gaussian+step distribution. The potentials 
correspond to $a=a_{cr}$ for each form of the scattering length. 
The continuous line shows the effective potential 
for $a=0$ (the linear trap). The frame at the left correspond 
to the emission of a single soliton in the case of a Gaussian+step 
modulation of $a$ (dotted curve). The shaded zone indicates the region
of negative scattering length in the step model.}
\label{fig3}
\end{figure}

The calculations, in the particular cases of $m=2$ --a Gaussian beam-- and 
$m\rightarrow\infty$ --a step function--, lead us to the effective potentials:
\begin{widetext}
\begin{subequations}
\begin{eqnarray}
\label{potgauss}
\Pi_g(z_0) & = & r_\perp^2\left[\frac{V_0}{\hbar\nu_\perp}
\left(1- \frac{e^{-\frac{z_0^2}{w^2+L^2}}}{\sqrt{1+\frac{w^2}{L^2}}}\right)+
\frac{aN}{w\sqrt{2\pi}} \left\{\text{erfc}\left[\frac{\sqrt{2}(D-z_0)}{w}\right]+ \frac{w_g}{\sqrt{w_g^2+w^2/2}}\text{exp}
\left[-\frac{(D-z_0)^2}{w_g^2+w^2/2}\right]\right.\right. \nonumber \\
& & \left.\left.\left[\text{erf}\left(\frac{-2z_0w_g+2w_gD+2w_g^2+w^2}{w\sqrt{2w_g^2+w^2}}\right)-\text{erf}\left(\frac{-2z_0w_g^2-Dw^2}{w_gw\sqrt{2w_g^2+w^2}}\right)\right]\right\}\right],\\
\label{potstep} \Pi_s(z_0)  & = &
 \frac{\sqrt{\pi}}{2}r_\perp^2
 \left\{\frac{V_0}{\hbar\nu_\perp}
\left(1- \frac{e^{-\frac{z_0^2}{w^2+L^2}}}{\sqrt{1+\frac{w^2}{L^2}}}\right) +
\frac{1}{\sqrt{2\pi}} \frac{aN}{w} \text{erfc}\left[\frac{\sqrt{2}\left(D-z_0\right)}{w}\right]\right\}, 
\end{eqnarray}
\end{subequations}
where $g$ and $s$ refer respectively to the step and Gaussian 
ramps. The functions $\text{erfc}(u)$ and $\text{erf}(u)=\frac{2}{\sqrt{\pi}}\int_0^u \exp(-v^2)dv=1-\text{erfc}$ 
are  the complementary error and the error 
functions, respectively. Fig. \ref{fig3} shows the equivalent potentials $\Pi$ 
given by Eq. (\ref{potgauss}) and Eq. (\ref{potstep}) for different values 
of $a$. As the scattering length becomes more negative, it reaches 
a limiting value $a_{cr}$ for which the potential $\Pi(0)=\Pi(\infty)$, 
thus if the atom cloud is initially placed at $z_0 = 0$ it will oscillate 
around the minimum 
and escape $z_0(\tau) \rightarrow \infty$ for $\tau \rightarrow \infty$, 
a phenomenon which is called soliton emission \cite{emision}. 
The critical value of $a$ that corresponds to the threshold for soliton 
emission  can be obtained within our formalism from the condition, 
$\Pi(0) = \Pi(\infty)$ which leads to:
\begin{subequations}
\begin{eqnarray}\label{acr}
Na_{cr}^g &=&  \frac{V_0w\sqrt{2\pi}}{\hbar\nu_\perp\sqrt{1+\frac{w^2}{L^2}}}\left[
\left\{\frac{1}{\sqrt{2}}\text{erfc}\left(\frac{D+w_g}{w}\sqrt{2}\right)+
\frac{w_g}{\sqrt{2w_g^2+w^2}}\text{exp}\left(\frac{-2D^2}{2w_g^2+w^2}\right)\right. \right.\nonumber\\
& &\left. \left. \left[\text{erf}\left(\frac{-2w_gD+2w_g^2+w^2}{w\sqrt{2w_g^2+w^2}}\right)- \text{erf}\left(\frac{-Dw}{w_g\sqrt{2w_g^2+w^2}}\right)\right]\right\}-\frac{2}{\sqrt{2}}\right]^{-1}, \\
Na_{cr}^s & = & \frac{\sqrt{2\pi}V_0}{\hbar\nu_\perp}\frac{wL}{\sqrt{L^2+w^2}}
\left[\text{erfc}\left(\frac{\sqrt{2}D}{w}\right)-2\right]^{-1}. 
\end{eqnarray}
\end{subequations}
\end{widetext}

In both cases, it is possible to find approximate expressions that 
hold for $D \approx w$ and $V_0=\Delta\hbar\nu_\perp$, with $\Delta<1$, which 
provides a shallow trap. Thus, it is possible to write:
\begin{subequations}
\begin{eqnarray}\label{Ncr}
Na_{cr}^g & \approx & - 2.655 \frac{\Delta w L}{\sqrt{L^2+w^2}} ,\\ 
Na_{cr}^s & \approx & - 2.565\frac{\Delta w L}{\sqrt{L^2+w^2}}.
\end{eqnarray}
\end{subequations}
We have found a good agreement between the exact numerical value and the 
prediction from Eq. (\ref{Ncr}). The variational model is more accurate 
for Gaussian potentials. The main discrepancy we have found after a wide 
numerical exploration of the parameter space is a factor $\sqrt{2}$ that 
is mainly due to the choice of the ansatz.

\begin{figure}[htb]
{\centering \resizebox*{1\columnwidth}{!}{\includegraphics{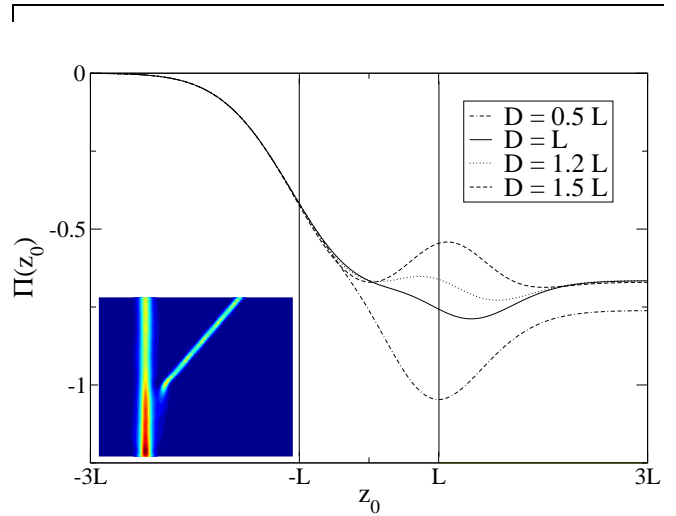}} \par}
\caption{[Color online] Same as Fig. \ref{fig3} for the Gaussian+step distribution.
The effective potentials correspond to different values of the penetration parameter $D$.}
\label{fig4}
\end{figure}

\section{ Partial outcoupling and multisoliton emission} 
\label{IV}

The distance $D$ between the edge of the region in which the scattering length 
varies and the center of the trap plays an important role in the emission process 
as shown in Fig. \ref{fig4}. As it can be seen in the plots, there is a 
critical overlapping between the $a<0$ region and the trap, for which the effective 
potential shows a central maximum between two adjacent minima. In this case, part of 
the cloud is reflected backwards by the intermediate barrier and thus, it is not 
possible to extract the whole cloud (see inset of Fig. \ref{fig4}). Thus, the emission 
will be of higher quality in traps of width below a certain critical threshold, 
which can be numerically calculated from the conditions $d\Pi/dz=0$ and $d^2\Pi/dz^2>0$.

\begin{figure}[htb]
{\centering \resizebox*{1\columnwidth}{!}{\includegraphics{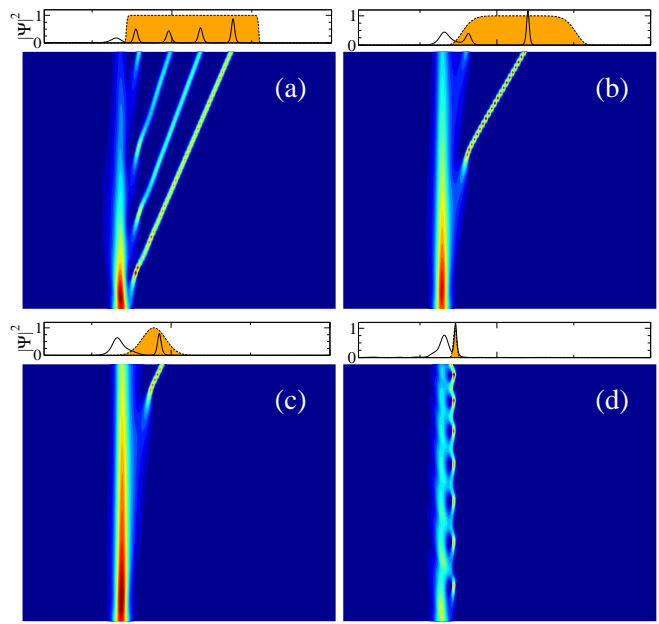}} \par}
\caption{[Color online] Emission of atomic solitons for different outcoupling
laser beams (shaded region in the top plots) of supergaussian shapes with different
values of the parameter $m$: a) $m=100$, b) $m=8$, c) and d) $m=2$ (Gaussian 
distributions of different widths). 
The small frames in the top of each picture display the profile of the solitons 
at $t=600\nu_\perp^{-1}$. Vertical axis in bottom pictures is time from 
$t=0$ to $t=600\nu_\perp^{-1}$. Horizontal axis is $60$ times the width $L$ 
of the gaussain trap. In all cases the product $Na$ is $8$ times the value of
Fig. \ref{fig4}. The rest of the parameters are the same as in Fig. \ref{fig3}.}
\label{fig5}
\end{figure}

A deeper numerical exploration based on Eq. (\ref{NLSE}) reveals
interesting effects beyond those contained in the averaged Lagrangian description. 
An example is shown in Fig. \ref{fig5}. As $a_*$ becomes more negative we obtain 
the emission of an integer number of solitons. 

In this section and to approximate our analysis to realistic scenarios we have considered the 
scattering length to be of the form
\begin{equation}
\label{gaus2}
a(z) = a \exp\left[-\left(\frac{z-D}{w_g}\right)^m\right],
\end{equation}
i.e., a supergaussian of finite width. 

\begin{figure}[htb]
{\centering \resizebox*{1\columnwidth}{!}{\includegraphics{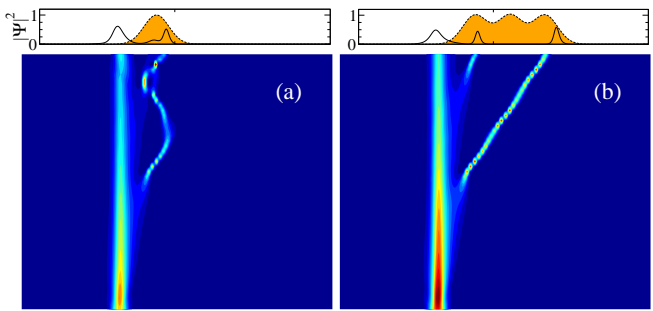}} \par}
\caption{[Color online] Emission obtained with the superposition of parallel
Gaussian beams. In (a) the  region of negative scattering 
length is generated by a single outcoupling beam as in Fig \ref{fig5} c). In 
(b) three equal Gaussian beams separated by its widths are used to generate a
wider region of negative scattering length. Top plots 
display the output at $t=1000\nu_\perp^{-1}$. In the pictures, vertical 
axis is time form $t=0$ to $t=1000\nu_\perp^{-1}$. The other 
parameters are the same as in Fig. \ref{fig5} c).}
\label{fig6}
\end{figure}

In Fig. \ref{fig5} we illustrate 
the dependence of the emission for different shapes of the function $a(z)$. We 
begin with a profile close to the step form ($m=100$) and then we relax the shape 
by diminishing the value of $m$ until reach the Gaussian function. We have also
observed that for finite $a<0$ regions, once the emitted solitons reach the opposite
edge of the zone with negative scattering length, they are reflected backwards and 
thus remain trapped in the vicinity of the BEC reservoir [see Figs. \ref{fig5}(d) and \ref{fig6} (a)]. 
This effect may have applications in the control of the emitted solitons 
once outcoupled and in the design of practical devices like laser tweezers for atoms \cite{tweezers}.

Another interesting consideration in order to construct experimentally 
wide regions of negative scattering length, is to use the superposition 
of several mutually incoherent laser beams as it shown in Fig. \ref{fig6}. 
In Fig. \ref{fig6} (a) the  region of negative scattering length is generated 
by a single outcoupling beam as in Fig \ref{fig5} c). As a result of the 
finite width of the outcoupling beam, the atomic soliton rebounds at
the boundary as is redirected backwards to the reservoir. In Fig. \ref{fig6} (b) 
three equal Gaussian beams separated by its widths are used to generate a wider 
region of negative scattering length. In the pictures, vertical axis is time 
form $t=0$ to $t=1000\nu_\perp^{-1}$. The other parameters are the same as 
in Fig. \ref{fig5} c).

We have also observed in the numerical simulations that the soliton emission 
is more efficient if the scattering length varies abruptly in space. 
Moreover, for avoiding the return of the solitons once they have been emitted, it is
desirable that the shape of the outcoupling beam be asymmetric. In the case in which 
laser beams are used to control the scattering length this could be 
achieved by partially overlapping parallel lasers. In order to generate a  extended
negative scattering length region, more lasers can be added or the outcoupling beam
can be displaced following the motion of the emitted soliton.

\section{Effect of temporal modulation of the output}
\label{V}
\begin{figure}[htb]
{\centering \resizebox*{1\columnwidth}{!}{\includegraphics{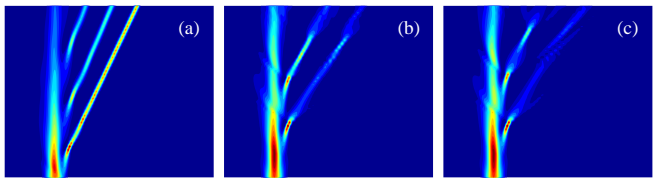}} \par}
\caption{[Color online] Effect of a temporal modulation in the emission process.
The spatial profile of the outcoupling beam is a step function that is 
modulated in time with a Gaussian ramp connected at its maximum to a flat top of variable 
size that ends with a final Gaussian decay. The captions correspond to different sizes of 
the plateau $t_p$ and separation between applied pulses $t_s$. In all cases time
goes from $t=0$ to $t=500\nu_\perp^{-1}$. a) corresponds to a continuous beam 
(i.e.: no modulation), in b) and c) we applied three pulses with 
$t_p=100\nu_\perp^{-1}$ in both cases and $t_s=5\nu_\perp^{-1}$ and 
$t_s=15\nu_\perp^{-1}$, respectively. The rest of the parameters are the same as
in \ref{fig5} a).
}
\label{fig7}
\end{figure}

Finally, we have also studied the temporal control of the emission process
by simulating the dynamics of the solitons outcoupled with pulsed beams.  
We have run calculations for different temporal sizes of the beams. The results
are shown in Fig. \ref{fig7}. In these simulations we have used temporal profiles
made with a Gaussian ramp connected to a flat top of variable size $t_p$. The spatial 
profiles are step functions. As it can be seen in the pictures, depending on the 
duration of the pulse (i.e.: the size of the flat top), the quality of the outcoupling 
varies, yielding to a broadening in the profile of 
the solitons as the nonlinearity vanishes. In some cases the solitons can fuse 
after the emission. By varying the time that $a$ is switched to a negative value, 
it is possible to obtain a periodic reconstruction of the individual 
solitons. The results are summarized in Fig.\ref{fig7}(c).In this simulation, 
the spatial profile of the outcoupling beam is a step function that is 
modulated in time with a Gaussian ramp connected at its maximum to a flat top of variable 
size that ends with a final Gaussian decay. The captions correspond to different sizes of 
the plateau $t_p$ and separation between applied pulses $t_s$. In all cases time
goes from $t=0$ to $t=500\nu_\perp^{-1}$. a) corresponds to a continuous beam 
(i.e.: no modulation), in b) and c) we applied three pulses with 
$t_p=100\nu_\perp^{-1}$ in both cases and $t_s=5\nu_\perp^{-1}$ and 
$t_s=15\nu_\perp^{-1}$, respectively. The rest of the parameters are the same as
in \ref{fig5} a).

\section{conclusions}

In summary, we have analyzed in detail the recently proposed mechanism for
outcoupling coherent matter wave pulses from a Bose-Einstein Condensate. By using this technique
it could be possible to obtain a regular and controllable emission of atomic soliton 
bursts that are easily extracted by an adequate choice of the control parameters. 

The particular shape of the spatial dependence of the scattering length used for 
outcoupling the solitons does not affect essentially the emission process although it is of higher
purity for sharp variations of the scattering length and for regions of negative 
scattering length of finite extent the soliton is reflected back when it reaches the boundaries. 
The temporal  control of the solitons can be easily implemented with pulsed beams.

We must stress that our results hold for different atomic
species like $^{85}$Rb and $^{133}$Cs, with adequate parameters. For the case of $^{133}$Cs
the scattering length can be controlled with high precision and can be 
done negative and large. This could provide a system in which a very 
high number of solitons could be generated.

Using the mechanism proposed in this paper a train of even several 
hundred of solitons could be coherently outcoupled from a condensate. As the techniques for coherently feeding 
the remaining condensate progress our idea could provide an outcoupling mechanism for a continuous 
atomic soliton laser.


We want to acknowledge J. Brand for discussions. 
This work was supported by Ministerio de Educaci\'on y Ciencia, Spain
(projects FIS2004-02466, BFM2003-02832 and network
FIS2004-20188-E), Xunta de Galicia (project PGIDIT04TIC383001PR) 
and Junta de Comunidades de Castilla-La Mancha (project PAI05-001).


\end{document}